\documentclass[ams,showpacs,nofootinbib,preprint]{revtex4-1}
\usepackage{amsmath}
\usepackage{epsfig}
\usepackage{color}
\usepackage{endnotes}
\usepackage{textcomp}
\usepackage{changes}

\newcommand{\be}{\begin{equation}}
\newcommand{\ee}{\end{equation}}

\newcommand{\ber}{\begin{eqnarray}}
\newcommand{\eer}{\end{eqnarray}}

\begin{document}

\title{Is the first excited state of the $\alpha$-particle a breathing mode? 
}

\author{Sonia Bacca$^{1,2}$,
  Nir Barnea$^{3}$,
  Winfried Leidemann$^{4,5}$, 
  Giuseppina Orlandini$^{4,5}$
  }

\affiliation{
  $^{1}$TRIUMF, 4004 Wesbrook Mall, Vancouver, B.C. V6J 2A3, Canada \\ 
  $^{2}$Department of Physics and Astronomy, University of Manitoba, Winnipeg, MB, R3T 2N2, Canada\\
  $^{3}$Racah Institute of Physics, The Hebrew University, 91904, Jerusalem, Israel\\
  $^{4}$Dipartimento di Fisica, Universit\`a di Trento, I-38123 Trento, Italy \\
  $^{5}$Istituto Nazionale di Fisica Nucleare, TIFPA, I-38123 Trento, Italy 
}

\begin{abstract}
The isoscalar monopole excitation of $^4$He is studied within a few-body {\it ab initio} approach. 
We consider the transition density to the low-lying and narrow $0^+$ resonance, as well as various sum rules
and the strength energy distribution 
itself at different momentum transfers $q$. Realistic nuclear forces of chiral and phenomenological nature are 
employed. Various indications for a collective breathing mode are found: i) the specific shape of the transition density,
ii) the high degree of exhaustion of the non-energy-weighted sum rule at low $q$
and iii) the complete dominance of the resonance peak in the excitation spectrum.
For the incompressibility $K$ of the $\alpha$-particle values between 20 and 30 MeV are found.
\end{abstract}

\bigskip


\maketitle

The quantum breathing mode (monopole oscillation) is the object of continuous theoretical and experimental investigations
in a large variety of systems as nuclei and trapped nanoplasmas or cold atoms (see e.g.~\cite{YoL13,AbB14} and references therein).
In fact it appears as one of the most important properties that allow to diagnose the underlying force.
The isoscalar giant monopole resonance (ISGMR), also known as nuclear breathing mode, is one of the collective nuclear
excitations that are well established and much discussed in heavier nuclei, 
also because of possible interesting relations to the nuclear matter incompressibility (see e.g.~\cite{KhM10,KhM12} and references therein).

The lightest nucleus where the breathing mode has been discussed is the $\alpha$-particle.
The discussion has a rather long history. It was triggered by the results of an inclusive electron scattering 
experiment on $^4$He in 1965~\cite{FrR65}, 
where  the transition form factor $|{\rm F}_{\cal M}(q)|^2$ to the  0$^+$ resonance (0$^+_R$) had 
been measured for various momentum transfer $q$. 
The interpretation of the resonance as a collective breathing mode  was suggested a year later~\cite{WeU66}.
A fair agreement both for the excitation energy $E_R$ and $|{\rm F}_{\cal M}(q)|^2$ was obtained.
In the following years the collectivity of the resonance was object of further discussion.

The intent of the present work is to analyze the collectivity issue of the 0$^+_R$ from a modern few-body 
{\it ab initio} point of view~\cite{LeO13}, as an interesting bridge between few- and  many-body physics. 
Further below it will become evident that various results of our {\it ab initio} few-body calculation
point to a breathing mode interpretation of the resonance.

As an introduction, we give a summary of the discussion on the 0$^+_R$, however leaving out 
other theoretical work  where the issue of collectivity is not addressed explicitly. 
After the work of~\cite{FrR65,WeU66} mentioned above, in 1970 the
0$^+_R$  was the object of an electron scattering experiment at lower $q$~\cite{Wa70}, and 
in the following decade the generator coordinate method~\cite{HiW53,Gr57} was applied 
to investigate the resonance~\cite{CaB73,AbC75}. 
Within this method collective motions are studied microscopically, introducing however a collective path. 
Again a fair agreement with experiment was found. The results gave  {\it ``further evidence that the first excited state in
$^4$He can be interpreted as a compressional monopole state''}~\cite{AbC75}. 
A translation invariant shell model calculation, which included two-particle two-hole (2p-2h)
configurations, was carried out in 1981~\cite{Sc81} and there it was concluded that the comparison of the obtained results 
with the experimental $|{\rm F}_{\cal M}(q)|^2$ {\it ``casts doubts on the usual breathing mode interpretation''}. 
A similar conclusion was drawn in 1986~\cite{LiZ86} by comparing  the results of a two- and four-particle
excitation model to data that included also new results at higher $q$~\cite{KoO83}. There it was stated that the 
resonance {\it ``has very little of the breathing mode in it and consists basically of more complex excitations''.}
In 1987 two  microscopic models were applied~\cite{VaR87}: one in terms of a collective variable, the other in 
terms of a cluster variable (similar to the resonating group approach).
It was  concluded that  the 0$^+_R$  {\it ``has a cluster character''} (3+1).
A year later it is stated~\cite{VaK88} that {\it ``these two hypotheses are not mutually exclusive''}.
In fact it was already shown in \cite{WeU66} that in a translation invariant harmonic oscillator
model, where one nucleon is excited from the $0s$ to the $1s$ shell, the transition density changes sign
at the $^4$He radius. This is precisely the form of the excess density of a  
breathing mode, though the  model may appear as non-collective, because of its interpretation as mean field or (3+1) cluster.
This common feature is  not very surprising in a  $s$-shell scenario, since the transition density for $q\to 0$
can be written as a function of just the hyperradius $\rho^2=\sum_i^4 r_i^2$, 
one of the six {\it collective} coordinates defined by the group $GL^{+}(3,R)$ 
~\cite{Zick71,Rowe80}.
At the end of the 1980s the 0$^+_R$  was understood as  {\it ``a superposition of simple 1p-1h excitations 
and not as a collective state''}~\cite{HaY89}.

We observe that in all these early work the criterion for the collectivity of the $0^+_R$  
has been mainly the agreement with experimental data:
if in a collective (non collective, i.e. purely mean field based) model the  data could be described sufficiently well,
it was simply concluded that the 0$^+_R$ has a collective (non-collective) character.

No further insight in the issue of collectivity came up till 2004, when it was
reconsidered within  a few-body {\it ab initio} approach~\cite{HiG04}.
Using a truncated version of a realistic  nucleon-nucleon (NN) potential, Argonne V8' (AV8')~\cite{PuP97}, 
and a phenomenological three-nucleon force (3NF), it was concluded with a sum rule argument that the $0^+_R$ 
is not a breathing mode. Another motivation against the collectivity was the large overlap 
of the $0^+_R$ wave function with the trinucleon ground state.  However, 
the density is an integral property, therefore it is perfectly possible, as already indicated in~\cite{WeU66,VaK88},
that a large overlap of the $0_R^{+}$ state with a (3+1) cluster
gives rise to the density of a breathing mode. The sum rule arguments are questioned below.

For many-body systems there is a general consensus that the breathing mode collectivity is signaled by two typical 
features:
(i) the above mentioned peculiar form of the transition density; 
(ii) the degree of exhaustion of the energy weighted sum rule~\cite{Fe57} by the resonance strength.
For $^4$He the latter was often examined, while the former was considered (besides in~\cite{WeU66}) only 
in ~\cite{HiG04}, where it was presented, but not commented. 

As already pointed out, the aim of the present work is the study of collectivity from a 
modern {\it ab initio} few-body  point of view. 
We will investigate whether an {\it ab initio} calculation of the $^4$He inelastic isoscalar monopole (InISM) strength   
exhibits features that are believed to characterize a collective
behavior in many-body systems, and how they depend on different nuclear forces. 
We will use two realistic potential models including the Coulomb force: 
the chiral effective field theory NN potential at next-to-next-to-next-to leading order (N$^3$LO)~\cite{EnM03} 
plus the N$^2$LO 3NF~\cite{Na07} and the AV18 NN potential~\cite{WiS95}
plus the UIX 3NF~\cite{PuP95} (both NN potentials lead to excellent 
fits to NN scattering data).
In this work we will consider both the form of the transition density to the  0$^+_R$ and the degree of exhaustion 
by the resonance strength of various SRs. 
We will investigate to what extent these features appear and conclude accordingly whether realistic  nuclear forces, 
the only input of a four-body {\it ab initio} calculation, allow\deleted{s}  to picture the excitation as  collective. 
In addition we show that the strength distribution itself leads to much better insights.

{\it Formalism.} 
The InISM strength distribution is given by 
\begin{equation}
   S_{\cal M}(q,\omega)= \sum_{n=0}^\infty\langle  n| {\cal M}(q)|0\rangle |^2\delta(\omega-E_n+E_0)\,,
\end{equation}
where $\omega$ is the excitation energy, and $|0\rangle ,|n\rangle $ and $E_0,E_n$ are eigenfunctions and eigenvalues 
of the nuclear Hamiltonian $H$, respectively  (for continuum energies the sum is replaced by an integral). 
The InISM operator reads 
\begin{equation}\label{bessel}
 {\cal M}(q)=\frac{1}{2}\left(\sum_{i=1}^A j_0(q r_i) -\langle 0|\sum_{i=1}^A j_0(q r_i)|0\rangle\right)\,,
\end{equation}
where $j_0$ is the zero-th order spherical Bessel function 
(the dependence on the nucleon form factor is neglected).

In the low-$q$, long wavelength (LW), limit the InISM operator is proportional to 
\begin{equation}\label{LW}
 {\cal M}^{\rm LW}=\frac{1}{2}\left(\sum_{i=1}^A  r_i^2 - A \langle r^2 \rangle\right)\,,
\end{equation}
where $\langle r^2\rangle$ is the mean square radius of the system. 
It is important to notice that ${\cal M}^{\rm LW}$\deleted{,}
depends only on the collective variable $\rho^2$, different from  ${\cal M}(q)$.

{\it Sum rules.} As well explained in~\cite{Ro70}, sum rules
{\it ``provide useful yardsticks for measuring quantitatively the degree of collectivity of a given excited state''}.
They are particular expressions for the moments~\cite{BoL79,LiS89,OrT91} defined as
\begin{equation}
  m_n(q)= \int d\omega \, \omega^n\,S_{\cal M}(q,\omega)\,.
\end{equation}
The moment $m_1$  for the operator ${\cal M}^{\rm LW}$  is an interesting SR. 
Using the completeness property of the eigenstates of the Hamiltonian $H=T+V$, where $T$ and $V$ are the kinetic 
and potential energy operators, one finds
\begin{equation}\label{m_1}
  m_1= \frac{1}{2}\langle 0|\left[{\cal M}, \left[T+V,{\cal M} \right]\right]|0\rangle \equiv m_1(T) + m_1(V)\,.
\end{equation}
For local potentials $m_1$ coincides with $m_1(T)$ and  in the LW limit one obtains 
\begin{equation}\label{FerrellSR}
  m_1^{\rm LW}(T)= \frac{1}{2}\langle 0|\left[{\cal M}^{\rm LW},\left[T,{\cal M}^{\rm LW}\right]\right]|0\rangle= 
  \frac{A}{2M}  \langle r^2\rangle\,,
\end{equation}
where $M$ is the nucleon mass and $A$ the number of nucleons ($\hbar=c=1)$.
Equation~(\ref{FerrellSR}) is known as the Ferrell energy-weighted SR (FEWSR)~\cite{Fe57}.

For $m_0$ completeness leads to  
\begin{equation}\label{m0}
m_0^{\rm LW}= \langle0|{\cal M}^{\rm LW}{\cal M}^{\rm LW}|0\rangle\,.
\end{equation}
While the FEWSR is considered  {\it ``model independent''}, $m_0^{\rm LW}$ is not. In fact
the experimental value of $\langle r^2\rangle$  (corrected for the nucleon finite size) can 
lead to a good estimate of the FEWSR. On the contrary for 
the evaluation of $m_0^{\rm LW}$ one would
need  one- and two-body ground-state densities. Particularly the latter is  largely model dependent.

For the special case where all the transition strength is concentrated into one specific excited state 
the sum rules become very simple.    
Assuming that the breathing mode $|\rm BM\rangle$ is such a state, it is evident that it exhausts
100\% of the FEWSR \cite{WeU66}. In general, this assumption implies that for any $n$ one has 
$m_n=(E_{\rm BM}-E_0)^n|\langle {\rm BM}| {\cal M^{\rm LW}}|0\rangle|^2$ and therefore
this state exhausts all sum rules completely:
\begin{equation}\label{rn}
r_n^{\rm LW}=\frac{(E_{\rm BM}-E_0)^n|\langle {\rm BM}| {\cal M^{\rm LW}}|0\rangle |^2}{ m_n}=1. 
\end{equation}

We note in passing that in~\cite{WeU66} also a simple  relation between the {$|{\rm F}_{\cal M}(q)|^2$
and the derivative of the elastic form factor was derived and that the same relation is found in~\cite{KaF70}
via the so-called ``progenitor sum rule''~\cite{No71}, under the hypothesis of a unique collective state.

It is generally believed that the ratios $r_n$, and in particular $r_1$, are good quantities to infer the degree 
of collectivity of a state.
However, it is usually neglected that for $n=1$ (and higher) one emphasizes the high-energy strength contribution. 
In fact, even in presence of a pronounced collective state, a negligible higher energy strength could lead to a rather small 
$r_1$ and thus to the wrong conclusion. In~\cite{Ro70} it is made clear that the proper quantity to check is rather $r_0$. 
It is the  {\it ``model independence''} of the FEWSR and the difficulty in calculating $m_0$,
especially for the heavier nuclei, that has led to the general attitude to consider $r_1$. 
For $^4$He, however, we are able to evaluate $m_0^{\rm LW}$ accurately. 
The results for the moments $m_0$ and $m_1$ of $S_{\cal M}(q,\omega)$, reported in the following, are
 obtained via the Lanczos algorithm, analogously to what was done in~\cite{GaB06} for the dipole operator. 
We perform our calculations by diagonalizing $H$
on the hyperspherical harmonics (HH) basis up to sufficient convergence, 
using the effective interaction HH  method~\cite{BaL00,BaL01}.

We test the accuracy of our results using the AV8' potential~\cite{PuP97}. 
Different from the other models this potential is {\it almost} non-local 
(the only non-local term, the spin-orbit term, does not contribute to $m_1^{\rm LW}(V)$). 
Therefore we get an estimate of our numerical error via independent calculations of 
$m_1^{\rm LW}$ and of $\langle r^2\rangle$ in Eq.~(\ref{FerrellSR}). 
We find $m_1^{\rm LW}=$ 183.43 fm$^4$ MeV to be compared to 183.62 fm$^4$ MeV from the FEWSR. 

{\it Results and discussion.} We discuss our first criterion for collectivity, namely the specific shape of the transition 
density\footnote{
This definition would in principle involve an integral over the resonance region, 
but, as in other calculations in the literature for very narrow resonances, we approximate $0^+_R$ by a bound state.
} 
\begin{equation}
\rho_{tr}(r)=|\langle0^+_R|{\cal M}^{\rm LW}|0\rangle|^2\,.
\end{equation}
In Fig.~\ref{figure1} we show $\rho_{tr}(r)$. One sees that the first criterion is met quite well since
 $\rho_{tr}(r)$ changes sign at a distance 
approximately equal to the root mean square radius of about 1.46 fm, for both our potential models.
A similar behavior is found in~\cite{HiG04} 
(see Fig. 2 there). Therefore we may conclude that this feature is rather independent on the force.
\bigskip\bigskip
\begin{figure}[htb]
\centerline{\resizebox*{8.5cm}{6cm}{\includegraphics[angle=0]{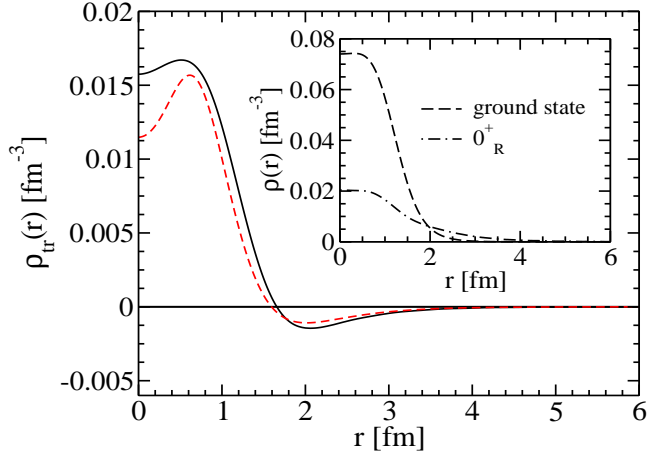}}}
\caption{(Color online) Transition density between the ground state 
and the 0$^+_R$ for the chiral (black full line) and the phenomenological (red dashed line) forces. 
In the inset the densities of the ground state and of the $0^+_R$ for the chiral force are also shown.
}
\label{figure1}
\end{figure}

Next we study our second criterion for collectivity, namely sum rules. 
First we analyze $r_0^{\rm LW}$, replacing in Eq.(\ref{rn}) $|{\rm BM}\rangle$ by 
$|0^+_R\rangle$.
We calculate the transition strength in the same way as we did in~\cite{BaB13},
but for ${\cal M}^{\rm LW}$. Here we recall that it is the Lorentz integral transform  method~\cite{EfL94,EfL07} 
and  a proper inversion algorithm that allows us to separate  the resonant from the background 
contribution.  For  the chiral (phenomenological) potential we find for the transition strength  
the value of 3.53 (2.25) fm$^4$ and   $m_0=6.80 \,(5.99)$  fm$^4$, 
leading to $r_0^{\rm LW}=52$\% (38\%). 
Considering the observation by Rowe~\cite{Ro70}: {\it ``A typical T=0 collective state exhausts something 
like 50 per cent''} (of  $m_0$), we are led to conclude 
that the chiral force generates a $0^+_R$ of {\it collective} character. 

Since in experiments the $|{\rm F}_{\cal M}(q)|^2$ is measured at finite momentum transfer, it is worth 
studying $r_n$ at various $q$ (in Eq.~(\ref{rn}): $r_n^{\rm LW}\to r_n(q)$, ${\cal M}^{\rm LW}\to{\cal M}(q)$).
\begin{table}
\caption{Transition form factor  $|{\rm F}_{\cal M}(q)|^2$  and
the zero-th and first moment of the strength distribution
for the InISM operator ${\cal M}(q)$. Also listed 
are the corresponding ratios $r_0$ and $r_1$. For each $q$ the upper and lower lines refer to  N$^3$LO+N$^2$LO
and AV18+UIX forces, respectively.}
\label{Table1}
\begin{center}
\begin{tabular}{c|c|cl|cc} \hline\hline
  $q$  &  {$|{\rm F}_{\cal M}(q)|^2$}  
               &  {$\,\,\,\,\,\,\,\,$\,\,$m_0\,\,\,\,\,\,\,\,$} 
               &  {$\,\,\,\,\,\,\,\,m_1\,\,\,\,$}           
               &  {$\,\,\,\,r_0\,\,$}  
               &  {$\,\,\,\,r_1\,\,$}\\

               [$\frac{{\rm MeV}}{{\rm c}}$]  &    
               &  
               &  {$\,\,\,\,$[MeV]}           
               &  {\%}  
               &  {\%}\\
 \hline              
 50  & 0.00034 & 0.00063 & 0.021 & 53  & 34  \\
     & 0.00024 & 0.00064 & 0.018 & 38  & 28  \\
 100 & 0.0042  & 0.0085  & 0.262 & 50  & 34  \\
     & 0.0031  & 0.0086  & 0.258 & 37  & 25  \\
 200 & 0.0248  & 0.0683  & 2.42  & 36  & 22  \\
     & 0.0190  & 0.0710  & 2.48  & 27  & 16  \\
 300 & 0.0297  & 0.129   & 5.89  & 23  & 11  \\
     & 0.0242  & 0.139   & 6.33  & 17  &  8  \\   
 400 & 0.0154  & 0.126   & 8.43  & 12  &  4  \\
     & 0.0141  & 0.143   & 9.39  & 10  &  3  \\
\hline
\end{tabular}
\end{center}
\end{table}
Results are given in Table~\ref{Table1}.
For both potential models $r_0(q)$ and $r_1(q)$  decrease with growing  $q$,
but differ somewhat. The low-$q$ the results for $r_0$ are in line with those in the $q\to 0$ limit.
For the chiral interaction $r_0$ reaches remarkable values of more than 50\%. 
In case of a collective mode such a $q$ behavior of $r_0$  can be expected, 
since low $q$ correspond to large wavelengths, where the virtual photon {\it ``sees''} the nucleus as a whole. 
The table also shows that  $r_1$ is always  smaller than  $r_0$. This 
indicates a non negligible high-energy strength and also shows that the argumentation 
against the collectivity which relies  on $r_1$ is based on shaky ground.
\bigskip\bigskip
\begin{figure}[htb]
\centerline{\resizebox*{8.5cm}{6cm}{\includegraphics[angle=0]{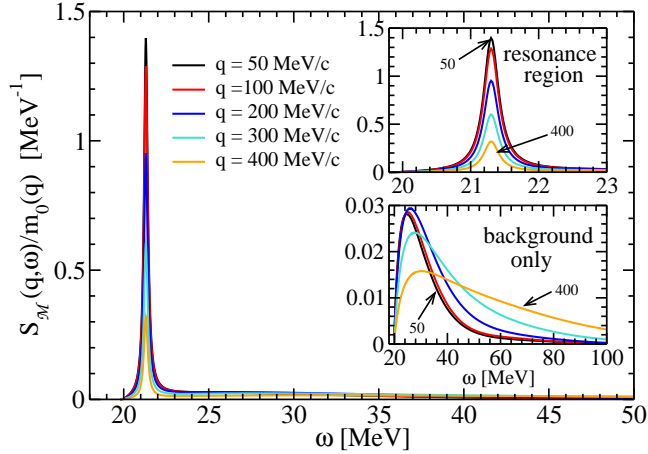}}}
\caption{(Color online) $S_{\cal M}(q,\omega)/m_0(q)$ for various fixed $q$. In the insets the strength in the resonance 
region and the background contribution.}
\label{figure2}
\end{figure}
Rather than  considering its integral properties the study of the strength distribution itself is much  more informative.
As already mentioned above, in~\cite{BaB13} we had computed both the InISM background distribution 
and the integrated $0^+_R$ strength (though not the strength distribution of the resonance itself). 
There we focused on $E_R$ and $|{\rm F}_{\cal M}(q)|^2$, which resulted to be largely
model dependent and considerably higher than existing data. 
Here, in Fig.~\ref{figure2} we show $S_{\cal M}(q,\omega)/m_0(q)$ at
constant $q$-values for the chiral interaction, assuming for the $0^+_R$ a Lorentzian with the 
experimental width of 270 keV~\cite{Wa70}. 
It is evident that the spectrum is completely dominated by the resonance peak. 
A similar result is obtained for the AV18+UIX potential even though the ratio resonance-background is somewhat
smaller. In Fig.~\ref{figure2} the peak becomes less pronounced with growing $q$, in favor of an increasing  
background (see insets of Fig.~\ref{figure2}). The more pronounced dominance at low $q$  can be understood if one considers
that ${\cal M}^{\rm LW}$ depends  only on the collective variable $\rho^2$.
This is not the case for ${\cal M}(q)$ in Eq.(\ref{bessel}), where  
the non-collective coordinates play a growing role with increasing $q$.

Here we would like to draw the attention to an interesting similarity between two very different physical systems:
the evolution with $q$ of the  monopole spectrum of this light system resembles that of the dynamical structure factor
of an alkali metal (almost free electron gas), where 
the plasmon (dipole) collective excitation is established at low $q$ (see e.g.~\cite{VoS89}).  

Now we turn to another aspect related to the breathing mode, namely the nuclear (in)-compressibility  and its relation to 
$m_{-1}^{\rm LW}$.
In~\cite{BoL79} it is shown that a monopole perturbation $V_P=\lambda\sum_i r_i^2$ induces a change of the radius
proportional to  $m_{-1}^{\rm LW}$ ($\lim_{\lambda\to 0}\delta\langle r^2\rangle/\lambda=-2\, m_{-1}^{\rm LW}$).
Therefore  $m_{-1}^{\rm LW}$ serves to define the nuclear incompressibility~\cite{Bl80}
$K_A^{I}=A \langle r^2\rangle^2/(2 m_{-1})$. We have calculated $m_{-1}^{\rm LW}$ 
summing the inverse-energy-weighted resonance strength
to the corresponding integral of the background contribution. For the chiral interaction and for AV18+UIX
we find  $m_{-1}^{\rm LW}=$ 0.259 and  0.236 fm$^4$ MeV$^{-1}$ (resonance strength contribution: 64\% and 45\%) and 
$\langle r^2\rangle =$ 2.146 and 2.051 fm$^2$, respectively. These values lead to $K_{4}^{I}=36$ MeV for both potentials.
Another definition of the nuclear incompressibility in terms of the resonance energy $K_A^{II}=E_R^2\, M \langle r^2\rangle$~\cite{Bl80}
is used in the literature. This is equivalent to $K_A^{I}$ if one uses the sum rule estimate 
$E_R=E_R^{SR} =\sqrt{m_1^{\rm FEWSR}/m_{-1}}$.
For N$^3$LO+N$^2$LO and AV18+UIX we find $E_R=$ 21.25 and 21.06 MeV, therefore $K_4^{II}=$ 23 and 22 MeV, respectively.
The reason why $K_4^{I}$ differs from $K_4^{II}$ is due to the discrepancy 
between the values of $E_R$ and the higher sum rule estimates $E_R^{SR}=$ 26.2 and 26.8 MeV, caused by 
the background contributions. All these values of the incompressibility are much smaller than the nuclear matter 
estimate ($230\pm 40$ MeV~\cite{KhM12}), showing the extreme softness of the $\alpha$-particle. 
Such low $K_{4}$ are in line with a recent parametrization  for nuclei with $A> 10$~\cite{VaT12}.
Extending the fit to $^4$He one obtains $K_{4}\simeq 0$, because of the large surface contribution.

{\it Summary.} We have investigated the InISM strength  of $^4$He 
and the corresponding  sum rules within a few-body {\it ab initio} approach, employing realistic nuclear forces 
(chiral and phenomenological ones).  For the $0^+_R$ we find properties that can also 
be attributed to a collective breathing mode: 
(i) the transition density changes sign at about the $^4$He radius; (ii) 
for low momentum transfer $q$, where the excitation operator can be expressed
in terms of the collective hyperradius, 
the transition strength to $0^+_R$ is large and exhausts between about 40\% (phenomenological force) 
and 50\% (chiral force) of the non energy-weighted sum rule and
(iii)  the resonance dominates the continuum spectrum completely over a very low and extended background. 

Moreover, we observe a very interesting similarity  in the evolution with $q$ of the spectrum, 
between our results  and the plasmon collective excitation  spectrum of an alkali metal. Finally, the inverse energy weighted SR
allows to give the incompressibility $K_A$ of $^4$He predicted by the modern realistic potentials: 22 $\leq K_A \leq$ 36  MeV.

A final clarification of the collectivity issue can only come from experiment.  Since the present criterion based on the FEWSR 
is not really appropriate, it would be 
necessary to determine the strength distribution at low $q$, where existing electron scattering 
data are scarce and limited to the resonance strength. Interesting complementary information 
could come from $\alpha$ scattering.

W.L and G.O. would like to thank E. Lipparini, S. Stringari and F. Pederiva for helpful
discussions. This work was supported in parts by the Natural Sciences
 and Engineering Research Council (NSERC), the National Research
 Council of Canada, the Israel Science Foundation (Grant number
 954/09), and the Pazi Fund.

\end{document}